\documentclass[fleqn,twoside]{article}
\usepackage{espcrc2}

\newcommand{\AmS}{{\protect\the\textfont2
  A\kern-.1667em\lower.5ex\hbox{M}\kern-.125emS}}

\title{Messages on Flavour Physics Beyond the Standard Model}

\author{Andrzej J. Buras \address[MCSD]{Physics Department, Technical
 University Munich
 \\  D-85748 Garching, Germany} }%

\begin{document}

\begin{abstract}
I present a brief summary of the main results on flavour physics beyond the
Standard Model that have been obtained in 2008 by my collaborators and myself
at the TUM. In particular I list main messages coming from our 
analyses of flavour and CP-violating processes in Supersymmetry, 
the Littlest Higgs model
with T-Parity and a warped extra dimension model with custodial protection
for flavour violating $Z$ boson couplings.

\end{abstract}

\maketitle

\section{Overture}

Elementary particle physicists are eagerly awaiting the first messages from
the LHC which, if we are lucky, will signal not only the discovery of the Higgs
 but also the existence of a definitive new physics beyond the
Standard Model (SM) of elementary interactions of quarks and leptons.

We need more than the SM in order to understand several observed facts, in
particular the huge hierarchy between the Planck scale and the electroweak
scale and the hierarchies in quark and lepton mass spectra and in their
flavour violating interactions summarized by the CKM and 
PMNS mixing matrices,
respectively. There are of course many other known reasons for going beyond
the SM and expecting  new physics at scales probed by the LHC but I will not
repeat them here.
While a large fraction of particle physicists bets that this new physics will
be supersymmetry, the turbulences on stock markets in this decade teach us
that it is wise to have many different shares.

 In this note I would like to report on the results of various analyses of
physics beyond the SM performed by my young and strong collaborators
 and
myself
 at the Technical University
Munich in 2008
 \cite{Buras:2008nn,BG2,Blanke:2008ac,Altmannshofer:2008hc,Altmannshofer:2008dz,Altmannshofer:2009ma,WED1,WED2,WED3}.

In view of space limitations this presentation will consist basically of a
list of messages that summarizes the main results of our papers on new 
physics, in particular supersymmetry, the Littlest Higgs model with T-parity
(LHT) and the Randall-Sundrum (RS) scenario with all particles 
except the Higgs 
propagating in the bulk.
All our papers deal with flavour violating processes, in particular
CP-violating ones. While in these papers we have hopefully cited properly
all relevant papers, the list of references presented here is 
incomplete and I apologize for it.
\boldmath
\section{$\varepsilon_{\rm K}$ -- An Old Star Strikes Back}
\unboldmath

One of the many successes of the SM is particularly striking. The SM is consistent 
within theoretical and parametric uncertainties
simultaneously with $|\varepsilon_{\rm K}|\approx 2.2\cdot 10^{-3}$
that measures tiny $K^0-\bar K^0$-mixing induced CP violation in
 $K_L\to \pi\pi$
decays and $S_{\psi K_{\rm S}}\approx 0.67$ that measures similar CP
violation in the $B_d^0-\bar B_d^0$ mixing. As $S_{\psi K_{\rm S}}$ is 
practically free of any non-perturbative uncertainties and 
$\varepsilon_{\rm K}$
involves the non-perturbative parameter $\hat B_{\rm K}$, it was
$S_{\psi K_{\rm S}}$ and not $\varepsilon_{\rm K}$ that together with 
the ratio of the $B_{d,s}-\bar B_{d,s}$ mass differences $\Delta M_d/\Delta M_s$
dominated unitarity triangle fits for the last five years.

This situation may change soon due to the following recent developments:

\begin{itemize}
\item
Improved lattice calculations of $\hat B_K$. In particular a recent
simulation  with dynamical fermions results in $\hat B_K=0.72\pm 0.04$
\cite{Antonio:2007pb}
that should be improved in the coming months,
\item
Inclusion of additional corrections to $\varepsilon_{\rm K}$
 \cite{Buras:2008nn}
 that were usually
neglected in the literature in view of the  $20\%$ error on 
$\hat B_{\rm K}$. As 
this parameter is now much better known it is mandatory to include them.
Effectively these new corrections can be summarized by an overall factor
in $\varepsilon_{\rm K}$: $\kappa_\varepsilon=0.92\pm 0.02$ \cite{Buras:2008nn}
\item
 As pointed out
in \cite{Buras:2008nn}   the decrease of $\hat B_{\rm K}$ relative to previous
 lattice results, that were in
the ballpark of $0.80$, together with $\kappa_\varepsilon$ being
 significantly below unity implies
within the SM  a
tension between the very precisely measured value of $S_{\psi K_{\rm S}}$ 
and $\varepsilon_{\rm K}$.
\end{itemize}

 Indeed $S_{\psi K_{\rm S}}=\sin 2\beta=0.671\pm0.024$
implies within the SM \cite{BG2}
\begin{equation}
|\varepsilon_K|^{\rm SM} = (1.78 \pm 0.25) \times 10^{-3}~,
\label{epsKSM}
\end{equation}
to be compared with
\begin{equation}
|\varepsilon_K|^{\rm exp} = (2.229 \pm 0.012) \times 10^{-3}~.
\label{epsKexp}
\end{equation}
If confirmed by a more precise value of $\hat B_K$ and 
more precise values of the CKM parameters, in particular $V_{cb}$, this would
signal new physics in $\varepsilon_{\rm K}$. Alternatively, no new physics in
$\varepsilon_{\rm K}$ would imply $\sin 2\beta=0.88\pm 0.11$ 
\cite{Lunghi:2008aa,Buras:2008nn}
 which could
only be made consistent with the measured value of
 $S_{\psi K_{\rm S}}$ by introducing a new phase $\phi_{\rm new}$
 in $B_d^0-\bar
  B_d^0$ mixing so that 
\begin{equation}
S_{\psi K_{\rm S}}=\sin(2\beta+2\phi^d_{\rm new})=0.671\pm 0.024
\end{equation}
with $\phi^d_{\rm new}\approx -9^\circ$.
Other
possibilities in which new physics would enter simultaneously in the $K$ 
and $B$ systems are discussed in \cite{Buras:2008nn,BG2}. Moreover it is
observed in \cite{Buras:2008nn} that a new phase 
$\phi^s_{\rm new}$ in  $B_s^0-\bar B_s^0$ mixing  
with $\phi^s_{\rm new}\approx \phi^d_{\rm new}$ would imply the sign and
the magnitude of the CP asymmetry $S_{\psi\phi}$ in accordance with 
the findings of CDF and D0 which  will be discussed below.

 Assuming that 
significant new physics contributions are present 
in $\varepsilon_{\rm K}$,
we explore in \cite{BG2} a number of
possibilities to  achieve an agreement with the experimental value of
$\varepsilon_K$.
In particular
we point out that within the CMFV framework (constrained minimal flavour 
violation)  this tension could be removed
with interesting implications for the allowed values of the $B$-meson decay
constants and/or branching ratios of rare $K$ decays. On the other
hand the MSSM with MFV and large $\tan\beta$ appears to worsen the situation.
A few observations are also made in the context of non-MFV new physics
scenarios.

\section{LHT Facing CP-Violation in $B_s^0-\hat B_s^0$ Mixing}
Another highlight of flavour physics in 2008 were the hints of a large new
CP phase in $B_s^0-\hat B_s^ 0$ mixing indicated by CDF and D0 data
\cite{Aaltonen:2007he,:2008fj,Brooijmans:2008nt}. They
imply the mixing induced CP asymmetry $S_{\psi\phi}$ in the ballpark of
 $0.4$, one order of
magnitude larger than its SM value: $(S_{\psi\phi})_{\rm SM}\approx 0.04$. 
Related studies by theorists can be found for instance in
\cite{Lenz:2006hd,Bona:2008jn,Faller:2008gt,Lenz:2008dp}.
If this large
value is confirmed with a small error, we will have a clear signal of
new CP-violating interactions beyond  CKM, falsifying thereby the concept of 
minimal flavour violation (MFV).

  A prominent model that goes beyond MFV is the LHT model  in which
the interactions between the SM quarks and the new heavy mirror quarks
mediated by new heavy weak gauge bosons involve a mixing matrix different from
the CKM matrix and consequently a new source of flavour and CP violation.
Already in 2006 we have pointed out that in this model $S_{\psi\phi}$ as large
as $0.3$ could be obtained \cite{LHT1}. Our updated 2008 analysis 
\cite{Blanke:2008ac} shows that
$S_{\psi\phi}$ can easily reach in the LHT model values $0.15-0.20$, 
while higher
values are rather unlikely though not excluded.

 Large enhacements are also possible in the branching ratios for 
$K_{\rm L}\to\pi^0\nu\bar\nu$, $K^+\to \pi^+\nu\bar\nu$ and 
$K_{\rm L}\to\pi^0 l^+l^-$ 
\cite{Blanke:2006eb,Goto:2008fj}
with much more modest effects in $B_{s,d}\to \mu^+\mu^-$ \cite{Blanke:2006eb}.
Finally the tension between $\varepsilon_{\rm K}$ and $S_{\psi K_{\rm S}}$ 
mentioned above can easily be resolved in this model.

\section{Low Energy Probes of CP Violation in a Flavour Blind MSSM}
All tensions between the SM and the data mentioned above can be removed in a
general MSSM with new flavour violating interactions coming predominantly from
the soft sector. However, the large number of parameters in this framework
does not allow for clear--cut conclusions. On the other hand the MSSM with MFV is
already rather constrained and CP violation and flavour violation being
governed solely by the CKM matrix  in MSSM-MFV are SM--like.

In this context an interesting alternative is the flavour blind MSSM (FBMSSM)
\cite{Baek:1998yn,Baek:1999qy,Bartl:2001wc,Ellis:2007kb,Altmannshofer:2008hc}
in which CKM remains to be the only source of flavour violation but new
CP-violating and {\it flavour conserving} phases are present in the soft sector. 
This
new physics scenario is characterized by a number of new parameters that is
 much
smaller than encountered in a general MSSM (GMSSM)
 and RS models 
discussed below. This implies striking correlations between
various observables that can confirm or exclude this scenario in the coming
years.

The main actors in the analysis of the FBMSSM in \cite{Altmannshofer:2008hc}
   are the CP asymmetry
 $S_{\phi K_{\rm S}}$ that experimentally is found  below its SM value,
electric dipole moments  of neutron and electron, the direct CP asymmetry 
$A_{\rm CP}(b\to s\gamma)$ and again $\varepsilon_{\rm K}$.

We find that $S_{\phi K_{\rm S}}$ can be made consistent with the
  experimental data but this requires sufficiently large values of the new
flavour conserving phases and automatically implies:
\begin{itemize}
\item
lower bounds on the electron and neutron electric dipole moments 
$d_{e,n}\ge 10^{-28}{\rm e~cm}$,
\item
positive and sizable (non-standard) $A_{\rm CP}(b\to s\gamma)$ in the
ballpark of $2\%-6\%$, that is roughly by an order of magnitude larger
than its SM value.
\item
under very mild assumptions also the $(g-2)_\mu$ anomaly can be explained.
\end{itemize}

On the other hand $S_{\phi\psi}$,  $S_{\psi K_{\rm S}}$ and
$\Delta M_d/\Delta M_s$ remain SM like but $|\varepsilon_{\rm K}|$ turns
out to be uniquely enhanced over its SM value up to a level of $15\%$. This
is certainly welcome in view of the discussion in Section 2. Clearly, it will
be very exciting to monitor the upcoming LHC, LHCb, Belle upgrade and
eventually Super-B factory in this and in the next decade to see whether
this simple and predictive framework can be made consistent with the data.

\boldmath
\section{A Goldmine of Observables: $B\to K^*\mu^+\mu^-$}
\unboldmath
In the difficult times at financial markets a goldmine is a very
useful thing to have. Such a goldmine is provided by the exclusive
decay $B\to K^*(\to K\pi)\mu^+\mu^-$ which will be studied in detail at LHCb.
Indeed various CP averaged symmetries and CP asymmetries resulting from
angular distributions offer 
24 observables which will provide an impressive amount of experimental 
numbers that will help to distinguish between various NP scenarios.
Model independent analyses of 
\cite{Bobeth:2008ij,Egede:2008uy,Hurth:2008jc}
have recently been generalized 
in \cite{Altmannshofer:2008dz},
where also specific models like MFV, the MSSM with MFV, the FBMSSM, the LHT 
and the GMSSM
have been analyzed. Moreover a number of correlations have been identified.
Several of the  CP averaged observables discussed in 
\cite{Altmannshofer:2008dz}
 can be considered as 
generalizations of the forward-backward asymmetry in $B\to K^*\mu^+\mu^-$. The
pattern of the zeros for these new asymmetries in
 a given model should be useful
in identifying the correct model or at least bound severely its parameters.
One of the important results of our studies is that new CP-violating 
phases will produce clean signals in CP-violating asymmetries.

Probably the most interesting results are found in the FBMSSM, in which
several symmetries and asymmetries differ significantly, even by orders
of magnitude, from the SM results and there exists a number of striking 
 correlations between these new observables and $A_{\rm CP}(b\to s\gamma)$
and $S_{\phi K_S}$. The NP effects in the LHT model are rather modest
except for a few CP asymmetries which are very strongly suppressed in
the SM. 

\boldmath
\section{Observables of $b\to s \nu\bar\nu$ Decays in the SM and Beyond}
\unboldmath
The rare decay $B\to K^*\nu \bar \nu$ is regarded as one of the important
channels in $B$ physics as it allows a transparent study of $Z$ penguin and 
other
electroweak penguin effects in NP scenarios in the absence of
dipole operator contributions and Higgs (scalar) penguin contributions that
are often more important than $Z$ contributions in $B\to K^*\mu^+\mu^-$ and
$B_s\to \mu^+\mu^-$ decays  \cite{Colangelo:1996ay,Buchalla:2000sk}.       
 In \cite{Altmannshofer:2009ma} we presented a new analysis of 
$B\to K^*\nu\bar\nu$ with improved formfactors and an update of 
$B\to K\nu\bar\nu$ and
 $B\to X_{s,d}\nu\bar\nu$
in the SM and in a number of NP scenarios. In particular various MSSM 
scenarios, 
the LHT model and a singlet scalar extension of the 
SM. The results for the SM and  NP scenarios can be transparently summarized in
 an $(\epsilon,\eta)$ plain analogous to the known $(\bar\varrho,\bar\eta)$ 
plane with a non-vanishing $\eta$ signalling this time 
not CP violation but the presence
of new right-handed down-quark flavour-violating couplings which can be
ideally probed by the decays in question.
Measuring the three branching ratios  and one additional
observable in $B\to K^*\nu \bar \nu$ allows to 
overconstrain the resulting point in the  $(\epsilon,\eta)$ plain with 
 $(\epsilon,\eta)=(1,0)$ corresponding to the SM. 
 We point out that  correlations with other rare decays 
 offer powerful tests of new physics with
new right-handed couplings and non-MFV interactions.

Concerning SM predictions, our result 
$Br(B\to K^*\nu\bar\nu)=(6.8\pm1.0)\cdot 10^{-6}$ is significantly lower
than the ones present in the literature. On the other our improved 
calculation of $Br(B\to X_s\nu\bar\nu)=(2.7\pm0.2)\cdot 10^{-5}$ avoids
the normalization to the $Br(B\to X_c e\bar\nu_e)$ and, with less than $10\%$
total uncertainty, is the most accurate to date.

\boldmath
\section{$\Delta F=2$ Observables in a Warped Extra Dimension with Custodial
  Protection Mechanism}
\unboldmath
Among  the most ambitious proposals to explain the hierarchy between the 
electroweak scale and the Planck scale \cite{Randall:1999ee} as well as of the observed hierarchical
pattern of fermion masses and mixings 
\cite{Gherghetta:2000qt,Chang:1999nh,Grossman:1999ra,Huber:2003tu,Agashe:2004cp} are models with
 a warped extra spatial
dimension, where the SM fields, except the Higgs boson, are allowed to 
propagate in the bulk. These models, called Randall-Sundrum (RS) models,
 provide a geometrical explanation of the
hierarchies in question.

Recently in a series of papers \cite{WED1,WED2,WED3}
 we have analyzed the electroweak and flavour
structure of a specific RS model based on the bulk gauge group
\begin{equation}
SU(3)_c\times SU(2)_L\times SU(2)_R \times U(1)_X \times P_{LR}.
\end{equation}
In this model the $T$ parameter \cite{Agashe:2003zs,Csaki:2003zu}
 and the coupling $Zb_L\bar b_L$
 \cite{Agashe:2006at} are protected
from new tree level contributions. This allows to satisfy the very 
precise electroweak constraints with  Kaluza-Klein (KK) masses
 of order 
$(2-3)~{\rm TeV}$ which are in the reach of the LHC.

Here I report first  the results of a complete study of 
 $\Delta S=2$ and $\Delta B=2$ processes in this model
 including $\varepsilon_{\rm K}$, $\Delta M_{K}$, 
 $\Delta M_s$, $\Delta M_d$,
 $A_{\rm SL}^q$, $\Delta\Gamma_q$, $A_{\rm CP}(B_d \rightarrow \psi K_S)$ and
 $A_{\rm CP}(B_s \rightarrow \psi \phi)$ \cite{WED1}. 
These processes are affected in this
 model  by
 tree level contributions from Kaluza-Klein gluons 
\cite{Agashe:2004cp,Burdman:2003nt,Csaki:2008zd} 
and new heavy electroweak
 gauge bosons $Z_H$ and $Z'$ \cite{WED1}, with
 the latter implied by the custodial protection mechanism.

It is in fact the first analysis in an RS model with this gauge
 group  that
\begin{itemize}
\item
simultaneously considers all the observables listed above,
\item
performs the full renormalization group analysis including the full basis of 
operators $Q_{\rm VLL}$, $Q_{\rm VRR}$, $Q_{\rm LR}^ 1$ and $Q_{\rm LR}^2$.
\item
in addition to tree level KK gluon exchanges considered in particular in 
\cite{Csaki:2008zd},
includes tree level contributions  of the two heavy weak gauge bosons $Z_H$
and $Z^\prime$ and of the KK photon $A^{(1)}$.
\end{itemize}

It is well known by now that in this framework explaining  the
hierarchies of fermion masses and weak mixing angles through different
positions of fermions in the bulk, necessarily leads to non-universalities
of gauge couplings to fermions. This in turn implies FCNC transitions at tree
level mediated by all neutral gauge bosons present in a given RS 
 model including
the SM $Z$ boson.

In the case of $\Delta S=2$ transitions  most dangerous are KK-gluon
exchanges that in the case of $\varepsilon_{\rm K}$ lead typically to 
values of this parameter by one to two orders of magnitude larger than
the measured value \cite{Csaki:2008zd}. Our more detailed analysis in 
\cite{WED1} that 
includes all relevant contributions confirms 
these findings:
  the fully anarchic approach to Yukawa couplings
 where all the hierarchies in quark masses
 and weak mixing angles are geometrically explained seems implausible and we 
confirm that the KK mass scale $M_{\rm KK}$ generically has to be at least
 {$\sim 20~{\rm TeV}$} to satisfy the $\varepsilon_{\rm K}$ constraint. 
We point out,
 however, that there exist regions in parameter space with only modest
 fine-tuning in the 5D Yukawa couplings which satisfy all existing
 $\Delta F=2$ and electroweak precision constraints for scales
 $M_{\rm KK}\simeq 3 ~{\rm TeV} $ in the reach of the LHC.  A recent detailed
analysis of $\varepsilon_{\rm K}$ in the so-called little RS model 
can be found in \cite{Bauer:2008xb}.

The additional specific new messages from our analysis are as follows
\cite{WED1}:

\begin{itemize}
\item 
 The EW tree level contributions to $\Delta F=2$ observables mediated by new
 $Z_H$ and $Z'$ weak gauge bosons, while subleading in the case of
 $\varepsilon_{\rm K}$ and $\Delta M_{\rm K}$, turn out to be of  
roughly the 
same size
 as the KK gluon contributions in the case of $B_{d,s}$ physics observables.
\item
 The contributions of KK {gauge boson} tree level exchanges involving new
 flavour and CP-violating interactions allow not only to satisfy all
 existing $\Delta F=2$ constraints but also to remove a number of
 tensions between  the SM and the data (see above), claimed in particular in
 $\varepsilon_{\rm K}$, $S_{\psi {\rm K_S}}$ and $S_{\psi \phi}$
 \cite{Buras:2008nn,Lunghi:2008aa,Lenz:2006hd,Bona:2008jn}.

\item Interestingly the model allows naturally for
 $S_{\psi \phi}$ as high as 0.4 that is hinted at by the most recent CDF and
  D{\O} data \cite{Aaltonen:2007he,:2008fj,Brooijmans:2008nt} and by an
 order of magnitude larger than the SM expectation: $S_{\psi \phi}\simeq
 0.04$.
\item
The tree level $Z$ contributions to $\Delta F=2$  processes are of higher
order in $v_{\rm SM}/M_{\rm KK}$ and can be neglected.
\item
We have pointed out that the 
 custodial symmetry $P_{LR}$ implies automatically the protection of
flavour violating
 $Zd_L^i\bar d_L^j$ couplings so that tree level $Z$ contributions to 
 all processes in which the flavour changes appear in the down quark sector
are  dominantly represented by $Zd_R^i\bar d_R^j$ couplings. This has profound
implications for rare $K$ and $B$ decays that we discuss  next.
\end{itemize}
A recent more detailed review
 of our results on $\Delta F=2$ appeared in \cite{Duling}.

\boldmath
\section{Rare $K$ and $B$ Decays in a Warped Extra Dimension with 
Custodial Protection}
\unboldmath
In \cite{WED2}
we present a complete study of rare $K$ and $B$ meson decays 
in the RS model with a custodial protection 
of left-handed $Z$ couplings just mentioned,
including   $K^+\to \pi^+\nu\bar\nu$, $K_L\to\pi^0\nu\bar\nu$, $K_L\to\pi^0 \ell^+\ell^-$,
$K_L\to \mu^+\mu^-$, $B_{s,d}\to \mu^+\mu^-$, $B\to K\nu\bar\nu$, 
$B\to K^*\nu\bar\nu$  and
$B\to X_{s,d}\nu\bar\nu$.  It turns out that new physics contributions to
these processes, 
as opposed to $\Delta F=2$ transitions, are governed
by
 tree level contributions from $Z$ boson exchanges 
(dominated by $Z d_R^i \bar d_R^j$ couplings) with the  KK 
 new heavy electroweak gauge bosons playing a subdominant role. Imposing
all existing constraints from $\Delta F=2$ transitions discussed above
and fitting all quark masses and  CKM mixing parameters we 
find that a number of branching ratios for rare $K$ decays 
can differ significantly from
the SM predictions, while the corresponding effects in rare $B$ decays 
are modest. In particular the branching ratios for $K_L\to\pi^0\nu\bar\nu$
and $K^+\to\pi^+\nu\bar\nu$ can be by a factor of 5 and 2 larger than the
SM predictions, respectively.
The latter enhancement could be welcomed one day if the central
  experimental value \cite{Artamonov:2008qb} will remain in the ballpark 
of $15\cdot 10^{-11}$ and its
error will decrease. However, it is very unlikely to get simultaneously large 
NP effects in rare $K$ decays and $S_{\psi\phi}$, which constitutes a
good test of the model.

We study correlations between 
various observables within the $K$ system, within the $B$ system and in 
particular between $K$ and $B$ systems.
For instance sizable departures from the MFV relations 
between $\Delta M_{s,d}$ and
    $Br(B_{s,d} \to \mu^+ \mu^-)$ and between $S_{\psi K_S}$ and the $K \to
    \pi \nu \bar \nu$ decay rates are possible. 
We also find that the pattern of
deviations from the SM differs from the deviations found in the
the LHT model \cite{Blanke:2006eb}.
 
 We  next show how
our results would change if the custodial protection of 
$Z d_L^i \bar d^j_L$ couplings was absent. As the custodial protection 
turns out to be more effective in $B$ decays, its removal implies a
possibility of large enhancements of rare $B$ decay branching ratios 
like $Br(B_s\to \mu^+\mu^-)$. On the other hand without this protection
it is  harder to obtain an agreement with electroweak precision 
data for KK scales in the reach of the LHC as summarized in 
\cite{Agashe:2003zs}. For a recent study of electroweak precision data
in a RS model without custodial protection see
\cite{Casagrande:2008hr}.

 It is interesting that in spite of many new flavour parameters present
 in this model a clear pattern of new flavour violating effects has been 
 identified
 in our analysis. Large effects in $\Delta F=2$ transitions, large effects
 in $\Delta F=1$ rare $K$ decays, small effects in $\Delta F=1$ rare $B$
 decays and the absence of simultaneous large effects in the $K$ and $B$
 system. This pattern implies that an observation of a large $S_{\psi\phi}$
 asymmetry would in the context of this model preclude sizable NP effects
 in rare $K$ and $B$ decays. On the other hand, finding $S_{\psi\phi}$ to 
 be SM-like will open the road to large NP effects in rare $K$ decays, even
 if such large effects are only a possibility and are not guaranteed.
 On the other hand, an observation of large NP effects in rare $B$ decays would
 put this model in serious difficulties. A recent more detailed review
 of our results appeared in \cite{Gori}.

\boldmath
\section{Electroweak and Flavour Structure of Warped 
 Extra Dimension Models with Custodial Protection}
\unboldmath
 Finally let me mention that the two phenomenological analyses in 
\cite{WED1,WED2} were based on a very detailed theoretical 
analysis \cite{WED3}, in which the electroweak and flavour structure of 
the model in question has been worked out.

Other selected recent studies of flavour violation in RS models
can be found in 
\cite{Agashe:2006iy,Cacciapaglia:2007fw,Fitzpatrick:2007sa,Cheung:2007bu,Csaki:2008eh,Santiago:2008vq,Agashe:2008uz}.
In particular in \cite{Agashe:2006iy,Agashe:2008uz} large contributions
to dipole operator dominated processes $\mu\to e\gamma$ and $B\to X_s\gamma$
have been found. On the other hand in
\cite{Cacciapaglia:2007fw,Fitzpatrick:2007sa,Cheung:2007bu,Csaki:2008eh,Santiago:2008vq} new strategies for the suppression of FCNC processes have been proposed.

\section{Final Messages}
As hopefully demonstrated above and in numerous papers on flavour physics in the
 literature, new exciting phenomena at scales of order $ 1~{\rm TeV}$
are possibly waiting for us. Let us hope we will meet them soon!

I would like to thank all my collaborators listed below for a wonderful
time we spent together exploring different avenues beyond the Standard
Model. I also thank Monika Blanke and Diego Guadagnoli 
for a critical reading of the manuscript.

Last but certainly not least I would like to thank Giulia Riccardi and 
her crew for a wonderful week I spent in Anacapri. I hope to be invited again!

\end{document}